\documentclass[aps,prl,amsmath,amsfonts,amssymb,twocolumn,superscriptaddress,showpacs]{revtex4-1}

\usepackage{graphicx,psfrag,color}
\usepackage{dcolumn}
\usepackage{bm}
\usepackage{mathbbol}

\newcommand{\med}[1]{\left\langle #1 \right\rangle}

\begin{document}

\title{Quantum correlations in spin chains at finite temperatures and quantum phase transitions}

\author{T. Werlang}
\affiliation{Departamento de F\'{i}sica,
Universidade Federal de S\~ao Carlos, S\~ao Carlos, SP 13565-905,
Brazil}
\author{C. Trippe}
\affiliation{Fachbereich C - Physik, Bergische Universit\"at Wuppertal, 42097 Wuppertal, Germany}
\author{G.A.P. Ribeiro}
\affiliation{Departamento de F\'{i}sica,
Universidade Federal de S\~ao Carlos, S\~ao Carlos, SP 13565-905,
Brazil}
\author{Gustavo Rigolin}
\email{rigolin@ufscar.br}
\affiliation{Departamento de F\'{i}sica,
Universidade Federal de S\~ao Carlos, S\~ao Carlos, SP 13565-905,
Brazil}

\date{\today}

\begin{abstract}
We compute the quantum correlation (quantum discord (QD)) and the
entanglement (EoF) between nearest neighbor qubits (spin-1/2)
in an infinite chain described by the Heisenberg model (XXZ Hamiltonian)
at finite temperatures. The chain is in the thermodynamic limit and
thermalized with a reservoir at temperature $T$ (canonical ensemble). We show
that QD, in contrast to EoF and other thermodynamic quantities,
spotlight the critical points associated to
quantum phase transitions (QPT) for this model even at finite $T$.
This remarkable property of QD may have important implications
for experimental characterization of QPTs when one is unable to reach
temperatures below which a QPT can be seen.
\end{abstract}

\pacs{03.67.-a, 03.67.Mn, 05.30.Rt}
\maketitle

Quantum phase transition (QPT) is a purely quantum process \cite{Sac99}
occurring at absolute zero temperature ($T=0$), where
no thermal fluctuations exist and hence no classical phase transition
is allowed to occur.
QPT is caused by changing
the system's Hamiltonian, such as an external magnetic
field or the coupling
constant.
These quantities are generally known as the tuning parameter.
As one changes the Hamiltonian one may reach
a special
point (critical point)
where the ground state of the system suffers
an abrupt change mapped to a macroscopic
change in the system's properties. This change of phase is
solely due to quantum fluctuations, which exist at $T=0$
due to the Heisenberg uncertainty principle.
This whole process is called
QPT. The
paramagnetic-ferromagnetic transition in some metals \cite{Row10},
the superconductor-insulator transition \cite{Gan10}, and
superfluid-Mott insulator transition \cite{Gre02} are remarkable
examples of this sort of phase transition.

In principle QPTs occur at $T=0$, which is unattainable
experimentally due to the third law of thermodynamics. Hence, one
must work at very small $T$, as close as possible to the
absolute zero, in order to detect a QPT. More precisely, one needs
to work at regimes in which thermal fluctuations are insufficient
to drive the system from its ground  to excited states.
In this scenario quantum fluctuations dominate and one is able to
measure a QPT.

So far the theoretical tools available to determine the critical
points (CP) for a given Hamiltonian assume $T=0$. For spin chains,
for instance, the CPs are determined studying, as one varies the
tuning parameter, the behavior of either its magnetization, or
bipartite \cite{Sar04} and multipartite \cite{Oli06} entanglement,
or its quantum correlation (QC) \cite{Dil08}. By investigating the
extremal values of these quantities as well as the behavior of
their first and second order derivatives one is able to spotlight
the CP. However, the $T=0$ assumption limits a direct connection
between these theoretical ``CP detectors'' and experiment. Indeed,
if thermal fluctuations are not small enough excited states become
relevant and the tools developed so far cannot be employed to
clearly indicate the CP.

In this Letter we remove this limitation and present a theoretical
tool that is able to clearly detect CPs for QPTs at finite $T$. We
show that the behavior of strictly QCs \cite{Zur01} at finite $T$,
as given by the thermal quantum discord (TQD) \cite{Wer10},
unambiguously detects CPs for QPTs that could only be seen, using
previous methods, at $T=0$ \cite{Dil08}. This remarkable property
of TQD, on one hand, is an important tool that can be readily
applied to reduce the experimental demands to determine CPs for
QPTs; or even allow such a detection for those systems where
today's technology makes it virtually impossible to achieve the
necessary $T$ below which quantum fluctuations dominate.  One the
other hand, this characteristic of TQD shows that QPTs have a
decisive influence on a system's physical property not only for
small $T$ but also above $T$ where quantum fluctuations no longer
dominate.

In order to show that TQD detects a QPT at finite $T$, we study
the anisotropic spin-1/2 Heisenberg chain (XXZ)
in the thermodynamic limit.
We assume the infinite chain to be in thermal equilibrium with a
reservoir at temperature $T$, i.e., its density matrix is
described by the canonical ensemble.
Tracing out all spins but the two nearest-neighbors we get their
reduced density matrix as a function of two-point correlation functions,
which are evaluated by solving a set of non-linear
integral equations (NLIE) \cite{Kluemper,NLIE}.
The two-qubit density matrix allows us to
compute TQD and investigate its properties
for $T>0$ as we change the system's
Hamiltonian. We show that TQD is maximal and its
first order derivative with
respect to the tuning parameter is discontinuous at the quantum
CP, not only at $T=0$ \cite{Dil08}, but also at
$T>0$. This behavior
is robust enough to be seen for high $T$.
Furthermore,
we have also computed the entropy, magnetization,
magnetic susceptibility, and
specific heat, for the whole chain, and two-site correlations
between the two nearest-neighbor spins
as well as their entanglement.
We show that none of these quantities detect unambiguously
the CP for $T>0$. We also discuss why TQD possesses
such a unique behavior in contrast to another
quantity, namely, the entanglement
between the two nearest-neighbors.

The XXZ Hamiltonian can be written as
\begin{equation}
H = J\sum_{j=1}^{L}\left(\sigma^{x}_{j}\sigma^{x}_{j+1} +
\sigma^{y}_{j}\sigma^{y}_{j+1} + \Delta
\sigma^{z}_{j}\sigma^{z}_{j+1}\right), \label{xxz}
\end{equation}
where periodic boundary conditions are assumed and $\Delta$ is the anisotropy parameter.
Here $L\rightarrow \infty$ and
$\sigma^{x}_{j}, \sigma^{y}_{j}$, and $\sigma^{z}_{j}$ are the usual
Pauli matrices acting on the $j$-th qubit. Throughout this Letter
$\hbar = 1$ and $J=1$ unless noted otherwise.
At $T=0$ the XXZ model has two CPs \cite{TAKAHASHI}.
At $\Delta=1$ we have a continuous phase transition
and at $\Delta=-1$ we have a first-order transition.

The density matrix for a system in equilibrium with
a thermal reservoir is $\rho = \exp{\left(
-\beta H  \right)}/Z$, where $\beta=1/kT$,  $Z = \mbox{Tr}\left\{ \exp{\left( -\beta H
\right)} \right\}$ is the partition function, and
the Boltzmann's constant $k$ is set to unity.
The nearest-neighbor two qubit state
is obtained by
tracing all but the first two spins,
$\rho_{12}=\mbox{Tr}_{L-2}\{\rho\}$.
Due to the translation invariance and $U(1)$
invariance ($[H,\sum_{j=1}^{L}\sigma^{z}_{j}]=0$)
of (\ref{xxz}),
we can write the reduced density matrix as follows,
\begin{equation}
\rho_{12}  = \left(
\begin{array}{cccc}
\frac{1+\med{\sigma_1^z\sigma_2^z}}{4} & 0 & 0 & 0\\
0 & \frac{1-\med{\sigma_1^z\sigma_2^z}}{4}  &
\frac{\med{\sigma_1^x\sigma_2^x}}{2} & 0 \\
0 & \frac{\med{\sigma_1^x\sigma_2^x}}{2} &
\frac{1-\med{\sigma_1^z\sigma_2^z}}{4} & 0 \\
0 & 0 & 0 &  \frac{1+\med{\sigma_1^z\sigma_2^z}}{4}\\
\end{array}
\right). \label{rho}
\end{equation}
These two-point correlation functions can be written
in its simplest form in terms of derivatives of the free energy
$f=(-1/\beta) \lim_{L\rightarrow \infty} (\ln{Z})/L$,
\begin{eqnarray}
\med{\sigma_j^{z}\sigma_{j+1}^{z}}=\partial_{\Delta}f/J,  \,\,\,\,\,
\med{\sigma_j^{x}\sigma_{j+1}^{x}}=(u-\Delta \partial_{\Delta}f)/2J,
\end{eqnarray}
with $u=\partial_{\beta} (\beta f)$ the internal energy.
In order to determine the free energy in the thermodynamic
limit and at finite $T$ one has to solve a suitable set of NLIE \cite{Kluemper,NLIE,footnote0}.

Now we can use (\ref{rho}) in order to show that the entanglement, as measured
by the entanglement of formation (EoF) \cite{Woo98}, is
%
$EoF$ $=$ $-g(f(C))$ $-$ $g(1-f(C))$,
%
with $f(C)=(1 + \sqrt{1 - C^2})/2$, $g(f)=f\log_2(f)$, and
%
\begin{equation}
 C = \mbox{Max}\{0,|\langle\sigma_1^x\sigma_2^x\rangle|-
|1+\langle\sigma_1^z\sigma_2^z\rangle|/2\} \label{C}
\end{equation}
the concurrence, an entanglement monotone. EoF quantifies a class
of QCs that cannot be created by local operations and classical
communication (LOCC) only \cite{Wer89}. Recently, however, it
became clear that there exist more general QCs if one removes the
LOCC restriction. These correlations are measured by the quantum
discord (QD) \cite{Zur01} and it is believed that QD quantifies
all correlations between two systems that has a pure quantum
origin. Note that EoF and QD coincide for bipartite pure states;
for mixed states, though, their difference becomes manifest being
both zero, however, when only classical correlations are present.
We can also conceptually understand QCs in comparison with
entanglement by noting that the latter is due to the superposition
principle applied to the whole Hilbert space of a bipartite
system. However, QCs as given by QD captures, on top of that, the
correlations coming from superposition of states within each
subsystem, a purely quantum effect that it is not possible
classically \cite{footnote1}. From this perspective, one can
better grasp why there exist states with zero entanglement but
finite QCs \cite{Ved10}. Another interesting and operational
interpretation for QD is achieved looking at the thermodynamic
properties of a quantum system. In \cite{Hor05} it is shown that
QD is related to the difference of work that can be extracted
acting either globally or locally at a heat bath with a bipartite
state when one-way communication is allowed.

For state (\ref{rho}) QD is \cite{Luo08}
$QD=$  $\left[g(1-2d_x-d_z)\right.$ $+$
$2$ $g(1+d_z)$ $+$ $\left.g(1+2d_x-d_z)\right]/4$
$-$ $\left[g(1+D)\right.$ $+$ $\left.g(1-D)\right]/2$,
with $d_x=\left\langle
\sigma_1^x\sigma_2^x\right\rangle$,
$d_z=\left\langle \sigma_1^z\sigma_2^z\right\rangle$, and
\begin{equation}
D =
\mbox{Max}\left\{\left|\left\langle
\sigma_1^x\sigma_2^x\right\rangle\right|,
\left|\left\langle \sigma_1^z\sigma_2^z\right\rangle\right|\right\}.
\label{D}
\end{equation}
Note that either $\left|\left\langle \sigma_1^x\sigma_2^x\right\rangle\right|$ or
$\left|\left\langle \sigma_1^z\sigma_2^z\right\rangle\right|$ is responsible for
the value of D.  As will be seen, it is the interplay between these two correlations
that is relevant in our understanding of why QD detects
a QPT at finite $T$ and EoF does not \cite{footnote2}.

\begin{figure}[!ht]
\includegraphics[width=8cm]{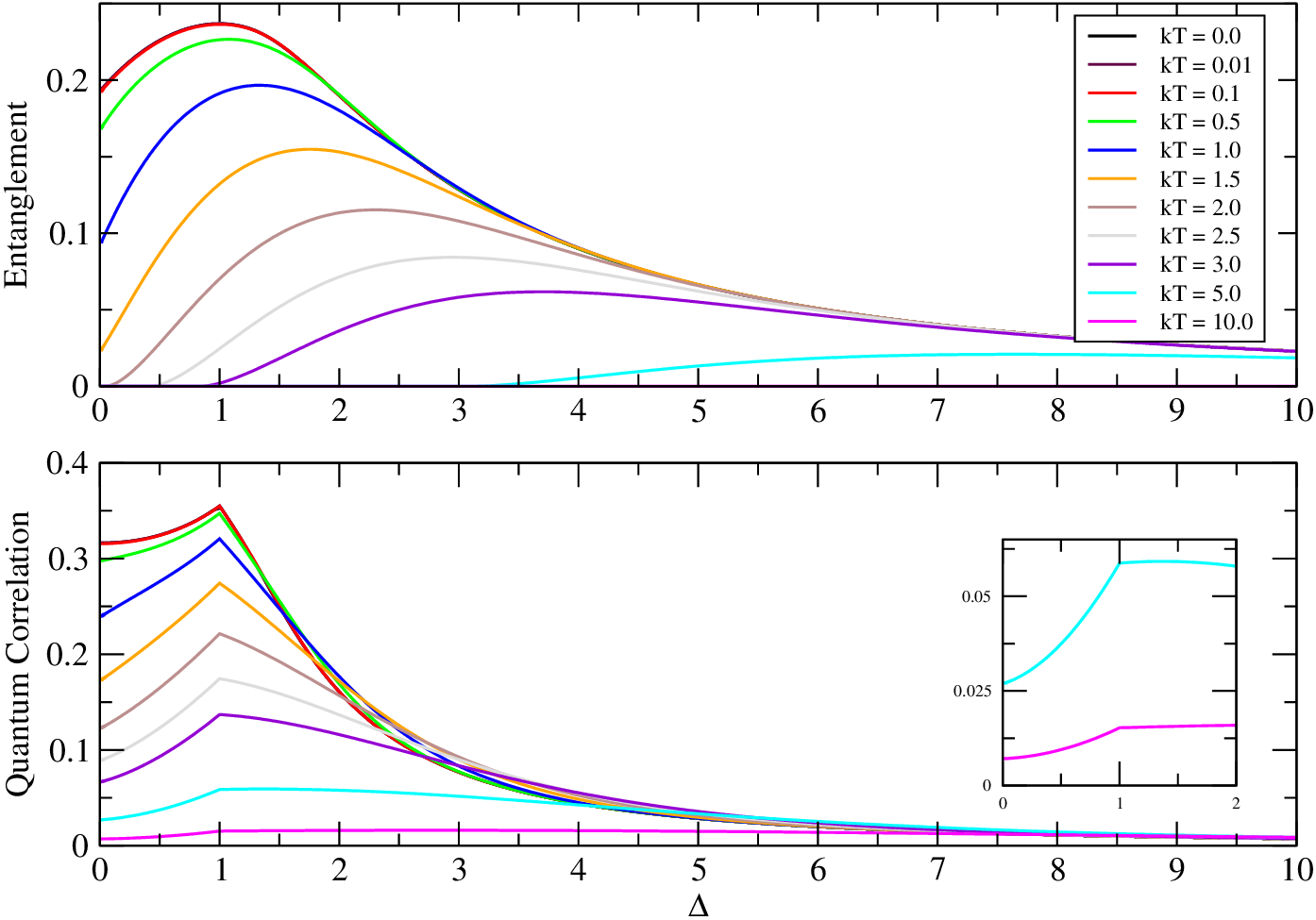}
\caption{\label{Fig1} (Color online) EoF (top) and QD (bottom) as functions
of the tuning parameter $\Delta$ for the XXZ model in the thermodynamic
limit. The inset depicts QD for high $T$ near the CP. $T$ increases from top to bottom.
The curves for $T=0$ and $T=0.01$ cannot be distinguished from the $T=0.1$.
Here and in the following graphics all quantities are dimensionless.}
\end{figure}

\begin{figure}[!ht]
\includegraphics[width=8.0cm]{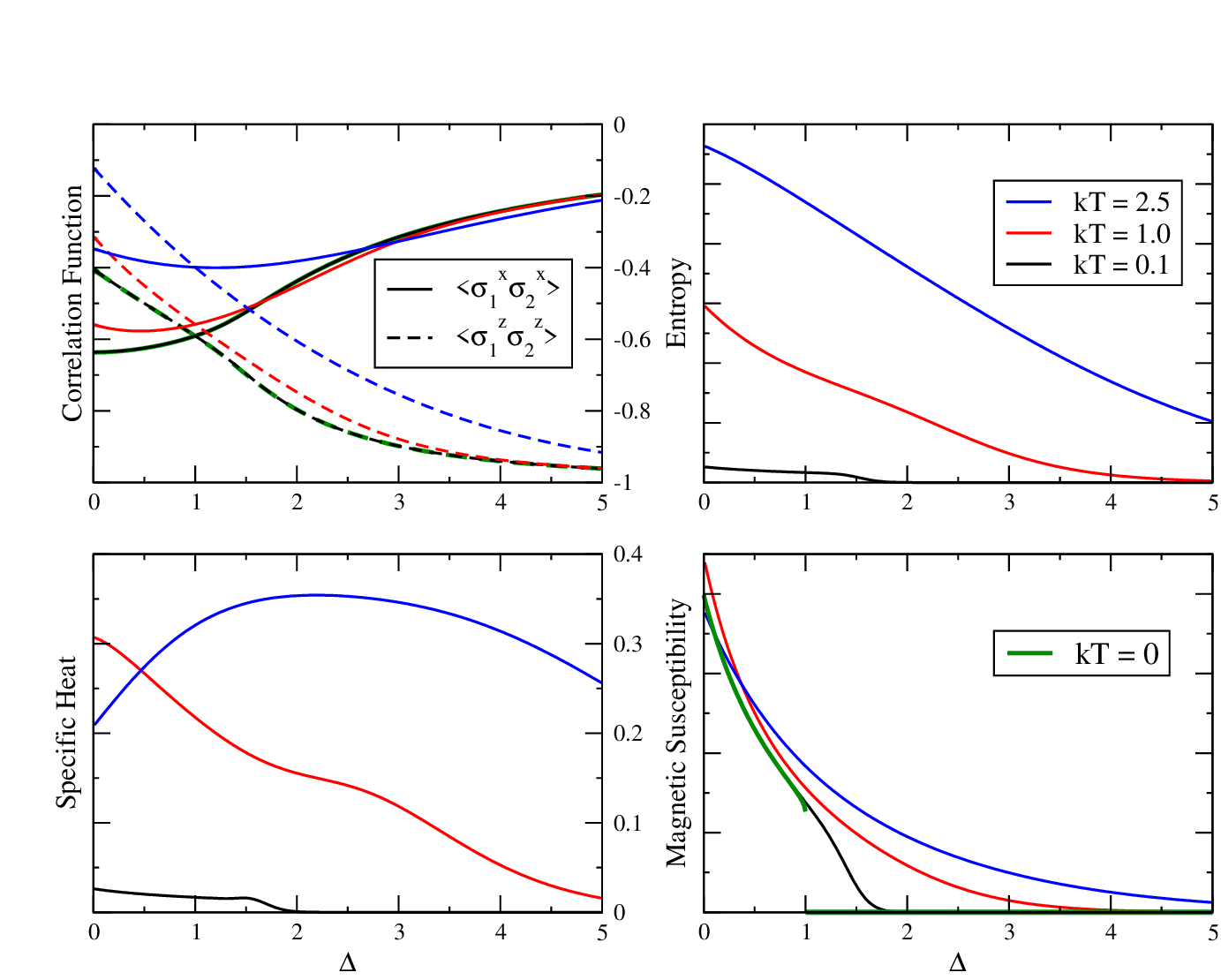}
\caption{\label{Fig2} (Color online) Thermodynamic quantities for the
XXZ model in the thermodynamic limit. The $T=0$ and $T=0.1$ curves for
the two-point correlation functions  are indistinguishable. Note that at $T=0$
the magnetic susceptibility also detects the phase transition being
discontinuous at the CP \cite{YANG}. The specific heat and entropy are null
at $T=0$.}
\end{figure}

We are now in a position to present the behavior of TQD and EoF
between two nearest-neighbor qubits
in an  infinite
spin chain at finite $T$. We first plot TQD and EoF, for several $T$,
as a function of the tuning parameter $\Delta$. This allows us
to prove the main claim in this Letter,
namely, that TQD detects a
CP of a QPT at finite $T$
while EoF does not. Looking at Fig. \ref{Fig1} we see that EoF is
maximal in the CP $\Delta = 1$ only at $T=0$, agreeing with the
results of \cite{Dil08}. As we increase $T$ the maximum no longer
occurs at $\Delta = 1$, moving to the region where $\Delta > 1$. Also,
the higher $T$ the farther from the CP is located the maximum of EoF.
On the other hand, TQD is maximal at $\Delta = 1$ when $T=0$ and
does not appreciably move away for $T\leq 3$. Moreover, its first
order derivative is discontinuous at the CP not only at $T = 0$
but also at $T>0$, a remarkable result showing that
TQD inherits at $T>0$ all of its important properties previously
seen only at $T=0$. This discontinuity of the first derivative of
TQD at $\Delta = 1$ is our CP detector for non null $T$.
In order to prove this unique behavior of TQD, we have
computed for several $T$ many thermodynamic quantities for the
infinite spin chain and also the pairwise correlations as a
function of the tuning parameter $\Delta$. As can be seen in Fig.
\ref{Fig2}, none of these quantities can clearly detect the CP at
$T>0$.

\begin{figure}[!ht]
\includegraphics[width=8cm]{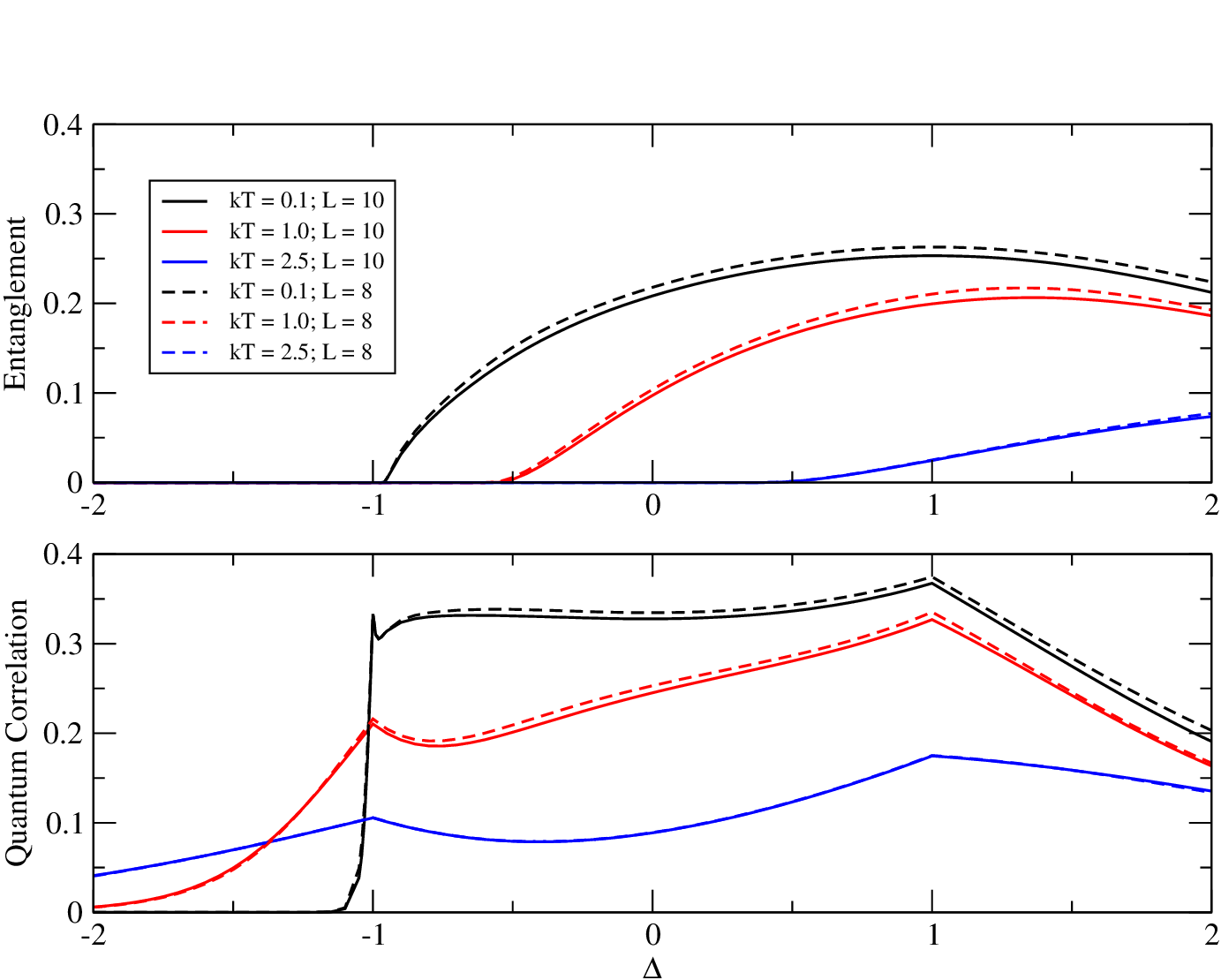}
\caption{\label{Fig3} (Color online) EoF and QD
for a chain of 8 and 10
qubits described by the XXZ model.
QD detects both quantum critical points at finite $T$ while
EoF does not.}
\end{figure}

Due to subtleties of the NLIE at $\Delta=-1$ ($J>0$), it is convenient to investigate 
how TQD behaves near the CP $\Delta = -1$ by means of 
numerical diagonalization of the Hamiltonian
(\ref{xxz}) \cite{footnote3}. We computed its thermal density matrix, and then
calculated the nearest-neighbor reduced density matrix for lattice
sizes $L=8$ and $10$.
Again, only TQD was able to
detect the CP for $T>0$. Looking at Fig. \ref{Fig3} we clearly see
that TQD successfully picks the CPs at $\Delta = \pm 1$ while EoF
does not. For finite $T$, the first derivative of TQD is
discontinuous at both CPs. EoF, on the other hand, is zero around
$\Delta = -1$ and its maximum gets shifted to the right at $\Delta
= 1$. Note that for small $T$ and $\Delta = -1$ TQD also resembles
its behavior at $T=0$, namely, being discontinuous at the CP
\cite{Dil08}.

\begin{figure}[!ht]
\includegraphics[width=8cm]{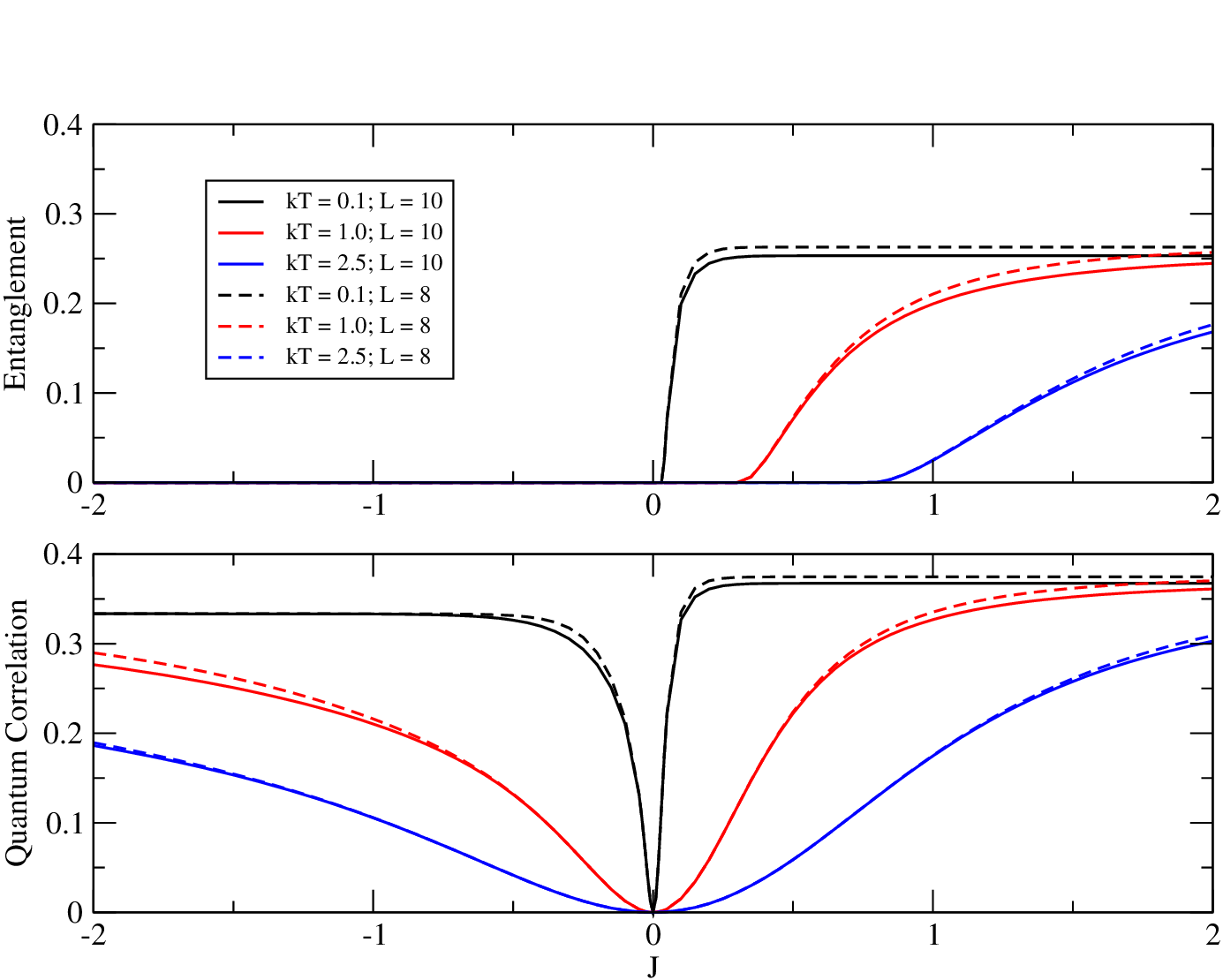}
\caption{\label{Fig4} (Color online) EoF and QD
for a chain of 8 and 10 qubits described by the XXX model.}
\end{figure}

In order to complement our results,
we fix the anisotropy parameter at $\Delta=1$ and then
vary the coupling constant $J$ from negative to positive values, i.e.,
we go from a ferromagnetic to an antiferromagnetic regime.
As can be seen in Fig. \ref{Fig4} TQD decreases
as one varies $J$ towards zero from both sides \cite{Wer10}.
Similar to the previous case, TQD inherits for finite $T$
its behavior at $T=0$. However, EoF is only non zero for the
antiferromagnetic regime; and for finite $T$ this only occurs away from
the vicinity of $J=0$. In other words, the behavior of TQD around $J=0$ and $T>0$
are qualitatively similar to its behavior at $T=0$ while this is no longer true
for the behavior of EoF.

We can understand this unique aspect of TQD, specially in contrast to
EoF, by taking a careful look at the analytical expressions giving EoF and TQD.
The main difference in behavior between EoF and TQD is connected to
Eqs. (\ref{C}) and (\ref{D}), being directly related to the maximization
process leading to these quantities. For the XXZ model and at finite $T$,
one can show that around the two CPs the function
maximizing (\ref{C}) does not abruptly change. It is either $0$ or
$|\langle\sigma_1^x\sigma_2^x\rangle|- |1+\langle\sigma_1^z\sigma_2^z\rangle|/2$.
On the other hand, for (\ref{D}), the function maximizing it changes exactly
at the CPs. Before the CPs one has either
$|\langle\sigma_1^z\sigma_2^z\rangle|$ or
$|\langle\sigma_1^x\sigma_2^x\rangle|$ as the maximum but after them
this role is exchanged. Indeed, in the vicinity of $\Delta<-1$
$D$ is given by $\left|\left\langle \sigma_1^z\sigma_2^z\right\rangle\right|$
while for
$-1 <\Delta < 1$
it is determined by
$\left|\left\langle \sigma_1^x\sigma_2^x\right\rangle\right|$
(see Fig. \ref{Fig2}).
Finally, in the vicinity of $\Delta>1$ it is determined by
$\left|\left\langle \sigma_1^z\sigma_2^z\right\rangle\right|$.
It is this change in the function maximizing $D$, which occurs
at $T=0$ \cite{Dil08} and shown here also to occur at $T>0$,
that is responsible for the discontinuity of
the first derivate of TQD. For the XXX model,
$|\langle\sigma_1^x\sigma_2^x\rangle|$ $=$ $|\langle\sigma_1^z\sigma_2^z\rangle|$,
and therefore no cusp-like behavior for TQD is observed. However,
TQD is only zero at $J=0$ for any $T$ while EoF is always zero in the vicinity of $J=0$
for $T>0$. Moreover,
working with small chains (up to 10 qubits) for various
$T$, we observed that the second derivative of TQD possesses a relatively high
value near $J=0$. We believe that it is likely that as one approaches
the thermodynamic limit the peak of the second derivative moves
towards $J=0$.

In summary, we presented a remarkable characteristic of quantum
correlations as given by the quantum discord (QD): its ability to
detect critical points (CP) of quantum phase transitions (QPT) at
finite $T$. Indeed, by solving an infinite chain described by the
XXZ model in the thermodynamic limit, we showed that QD is able to
highlight the CPs of QPTs for $T>0$ while neither the entanglement
nor any thermodynamic quantity achieve the same feat. This
property of QD may be useful in the experimental detection of CPs
for QPTs where one is not able to reach the temperatures below
which a QPT can be seen. Conceptually, this capacity of QD to
detect CPs of QPTs for $T>0$ and its interesting and puzzling
dynamical robustness against noise \cite{Maz10,Guo10} illustrate
the broad range of scenarios where QD helps in the understanding
of fundamental issues of quantum mechanics.

\begin{acknowledgments}
TW thanks CNPq for funding.
CT thanks Volkswagen Foundation for financial support.
GAPR thanks FAPESP for funding.
GR thanks CNPq/FAPESP for financial support through the
National Institute of Science and Technology for Quantum Information.
GAPR and CT thank Frank G\"ohmann for discussions.
\end{acknowledgments}


\begin{thebibliography}{99}
\bibitem{footnote0} Details of the calculations leading to the two-point
correlation functions and a complete analysis of this model with and without
magnetic field will be given elsewhere.
\bibitem{footnote1} The superposition principle
allows a subsystem of a bipartite system to be prepared in a
superposition of two non-orthogonal states. In such a case
the whole system cannot be described classically resulting in $QD\neq0$.
\bibitem{footnote2} Technically, which correlation
function dominates is related to the maximization, over all
possible measurements on one of the subsystems, carried out to
compute an extension of the classical mutual information (MI) for
quantum states using the Baye's rule. This quantity is then
subtracted from another way of defining MI without Baye's rule.
The difference between these two quantities leads to QD
\cite{Zur01}. For states as (\ref{rho}) the maximization leading
to QD is achieved by either one of two different sets of
measurements, defining which correlation function is the relevant
one in the expression for QD \cite{Wer10,Maz09}.
\bibitem{footnote3} It is worth noting that the two-point correlations for
$L=10$ are very close to those (Fig. \ref{Fig2}) at the
thermodynamic limit for $\Delta>0$.
\end{thebibliography}
\end{document}